\documentclass[a4paper,11pt]{article}
\usepackage{pos}

\usepackage{amsmath}
\usepackage{mathtools}

\title{Di-$\Jpsi$ structures from the quark Pauli-blocking effect}

\author*[a,b,e]{Sachiko Takeuchi}
\author[b]{Atsushi Hosaka}
\author[c,e,f]{Makoto Takizawa}
\author[d,g]{Yasuhiro Yamaguchi}

\affiliation[a]{Japan College of Social Work,
Tokyo 204-8555, Japan}
\affiliation[b]{Research Center for Nuclear Physics (RCNP), Osaka Univ., Ibaraki, Osaka, 567-0047, Japan,}
\affiliation[c]{Showa Pharmaceutical University, Higashi Tamagawa Gakuen 3-2-1, Machida, Tokyo 194-8543, Japan}
\affiliation[d]{Department of Physics, Nagoya University, Chikusa-ku Furocho, Nagoya, Aichi 464-8601, Japan}
\affiliation[e]{RIKEN Nishina Center, Wako, Saitama 351-0198, Japan}
\affiliation[f]{J-PARC Branch, KEK Theory Center, Shirakata 203-1, Tokai, Ibaraki 319-11006, Japan}
\affiliation[g]{Kobayashi-Maskawa Institute for the Origin of Particles and the Universe, Nagoya University, Nagoya,  Aichi 464-8602, Japan}

\emailAdd{sachitak@rcnp.osaka-u.ac.jp}

\abstract{%
The double-charmonium scattering states such as $\Jpsi\Jpsi$, $\eta_c\Jpsi$, and $\eta_c\eta_c$ are investigated
by a simplified quark cluster model.
It is found that the quark Pauli-principle over the $c\bar c c\bar c$ system 
causes a rapid increase and a node in the two-meson phase shifts.
The increase is not large enough to be regarded as a resonance, 
but if it is seen experimentally, that is most likely the quark Pauli-blocking effect.
}

\FullConference{The XVIth Quark Confinement and the Hadron Spectrum Conference (QCHSC24)\\
 19-24 August, 2024\\
 Cairns Convention Centre, Cairns, Queensland, Australia\\}

\def\bra{\langle}
\def\ket{\rangle}
\def\rmd{{\rm d}}

\def\calN{{\cal N}}

\def\pii[#1]{{1\over (2\pi)^{#1}}}

\newcommand{\xbld}[1]{\mbox{\boldmath $#1$}}

\def\vecr{\xbld{r}}

\def\Jpsi{{J\!/\psi}}
\def\Psfc{{P^{\rm sfc}_{24}}}
\def\Porb{{P^{\rm orb.}_{24}}}

\begin{document}
\maketitle

\section{Introduction}
In 2020-24, the observation of $c\bar c c\bar c$ exotic hadrons (di-$\Jpsi$) was reported by LHCb, ATLAS, and CMS
\cite{LHCb:2020bwg,ATLAS:2023bft,CMS:2023owd}. 
Their masses are above the $2m(\Jpsi)$ threshold, $X(6200, 6600, 6900)$. 
The Belle experiment also conducted investigations near the $\eta_c\Jpsi$ threshold; 
however, no evidence has been observed to date \cite{Belle:2023gln}.
Many theoretical works have been done to describe the states, but since color-spin interaction between quarks is repulsive or zero in the $S$-wave $c\bar cc\bar c$ states, 
and since there is no long-range attraction from the pion exchange, it seems difficult for the quark models to have resonances below 7GeV \cite{Santowsky:2021bhy,Ortega:2023pmr,Meng:2024yhu,Wang:2023jqs}.

In this work, we investigate $c\bar cc\bar c$ systems focusing on the quark-Pauli blocking effect. 
Pauli forbidden principle over the quarks definitely exists in these fermion composite systems but has not yet been focused on in the scattering states.
We employ the quark-hadron hybrid model that has a short-range four-quark part with a two-$c\bar c$-meson tail. 
The quark Pauli-principle ‘blocks’ the two quarks from occupying the same state when the two mesons come close.
As a result, one of the (anti)quarks moves from the $0s$- to the $1s$-level,
which can be written as a potential like
$v_0(|0s\ket\bra 1s|+|1s\ket\bra 0s|)$. 
As we will show later, this term arises from the kinetic term of the Hamiltonian when the quark degrees of freedom are introduced.
The obtained potential has a repulsion in the intermediate range and a strong attraction in the short range.
This potential of the well-and-barrier type gives rise 
a rapid increase of the phase shifts at around $E$ = 6.4 GeV. 
These effects indeed should be considered when one investigates the di-$\Jpsi$ spectrum.

\section{Model}

We use a simplified nonrelativistic quark model to investigate the $c\bar cc\bar c$ systems.
The $c$-quark and $\bar c$-antiquark are assumed to form meson clusters,
and antisymmetrized to take into account the quark Pauli-principle.
This arises the quark interchange between the two mesons.
As for the interaction between the quarks,
we only consider the color-spin interaction explicitly.
We assume that the other terms, such as the Coulomb and the confinement potentials, are used up 
to make the meson clusters, 
and that the short-range color- or color-spin interaction remains as a residual interaction.
This simplified model allows one to investigate the scattering behaviors of systems with the quark configuration more easily and discuss the situation almost free from parameter choice.

\subsection{Model space}

A $c\bar c$ meson with a spin $\alpha$ 
can be written as
\begin{align}
\phi_\alpha&=\phi_{0s}(\tilde b,\vecr_{12})c^\alpha_{\alpha_1\alpha_2}|c(\alpha_1)\bar c(\alpha_2)\ket~,
\label{eq:mesonwf}
\end{align}%
where $\phi_{0s}(\tilde b,\vecr)$ is the $0s$ harmonic oscillator wave function with a size parameter $\tilde b$,
$\vecr_{12}$ is the relative coordinate of the quark and the antiquark, 
$\alpha_i$'s are the spin and the color of the $i$th (anti)quark.
$\alpha$ and $c^\alpha$ are the spin of the meson and Clebsch-Gordan coefficient, respectively.
Since the color is taken to be singlet, 
 $\phi_1$ corresponds to $\Jpsi$ and $\phi_0$ to $\eta_c$.

For simplicity, we use a Gaussian wave function for the quark internal motion in the mesons.
Its size parameter $\tilde b$ is taken so that the root-mean-square radius of the meson is consistent with the observed $c\bar c$ excitation energy.
With $m(\psi(2S))-m(\Jpsi)\sim$600 MeV and with the $c$-quark mass of 1500 MeV 
$\tilde b$ becomes 0.385 fm \cite{Takeuchi:2024syc}.

As for the wave functions of the two-meson systems, we take
\begin{align}
\Psi_J &= (1-\Psfc\Porb)\psi_J\,\chi_J(r)~,
\\
\psi_J &= 
\sum_{\alpha_a\alpha_b}c^J_{\alpha_a\alpha_b}\,\phi_{\alpha_a}\phi_{\alpha_b}
=\phi_{\alpha_a}\phi_{\alpha_b}\big|_J~,
\end{align}%
where $J$ is the spin of the two-meson system,
and $\Psfc$ and $\Porb$ are the quark exchange operators
between the two $c$- or $\bar c$-quarks 
in the spin-flavor-color space and the orbital space. 
The orbital wave function, $\chi_J(r)$,
is taken to be $S$-wave and $r$ is the distance between the two meson clusters.
For the $c\bar c c\bar c$ systems in the orbital $(0s)^4$ configuration, 
there is one state for $J=2$,
one for $J=1$, and two for $J=0$. 
\begin{align}
%
\psi_{J=2} &= \phi_1\phi_1 \big|_{J=2}
\label{eq:psiJ2}
\\
%
\psi_{J=1} &= {1\over \sqrt{2}}\Big(\phi_0\phi_1 \big|_{J=1}+\phi_1\phi_0 \big|_{J=1}\Big)
\label{eq:psiJ1}
\\
\psi_{J=(00)0} &=\phi_0\phi_0\big|_{J=0} 
~~,~~~~
\psi_{J=(11)0} =\phi_1\phi_1\big|_{J=0}~~.
\label{eq:psiJ0b}
\end{align}%
The quark antisymmetrization conserves the total angular momentum, $J$, 
but mixes the two $J=0$ wave functions denoted as $\psi_{J=(00)0}$ and $\psi_{J=(11)0}$ above.


\subsection{Hadron Hamiltonian obtained from the quark model}

We employ the Hamiltonian for quark systems as
\begin{align}
H_q&=H_0 + V_{c\bar c}+V_{cc}+V_{\bar c\bar c}
\label{eq:Hamiltonian}
\\
H_0 &= \sum_{i}\Big(m_i+{p_i^2\over 2m_i}\Big) - {p_G^2\over 2m_G}
\\
V_{c\bar c}&=\sum_{i=c,j=\bar c} \Big(
{(\lambda_i\cdot\lambda_j})a_{c\bar c}
-(\lambda_i\cdot\lambda_j)(\sigma_i\cdot\sigma_j)c_{c\bar c}
\Big){\cal P}^{0s}
\label{eq:Vccbar}
\\
V_{cc,\bar c\bar c}&=-\sum_{ij=cc ~or~ \bar c\bar c} (\lambda_i\cdot\lambda_j)(\sigma_i\cdot\sigma_j) c_{c c}~{\cal P}^{0s}~,
\label{eq:Vcc}
\end{align}%
where $m_i$ and $p_i$ are the mass and the momentum of the $i$th (anti)quark, and
$m_G$ and $P_G$ are the total mass and the center of mass momentum, respectively.
The potential between the $c$ and $\bar c$ quarks, $V_{c\bar c}$, consists of the spin-independent color term, $(\lambda\lambda)$,
and the color-spin term, $(\lambda\lambda)(\sigma\sigma)$.
The potential between the two charm quarks or the two antiquarks
$V_{c c}$ and $V_{\bar c\bar c}$ also has $(\lambda\lambda)$ and $(\lambda\lambda)(\sigma\sigma)$ terms originally. 
However, we do not include the former here because the term vanishes when it interacts between the two color-singlet clusters of the same quark mass.
${\cal P}^{0s}$ is the projection operator onto the $(0s)^4$ configuration 
when the interacting quarks belong to the different mesons. When they belong to the same meson, it is just 1.
Namely, the quark interchange is assumed to occur between the two clusters only when the four quarks form the $(0s)^4$ configuration.

The present model does not depend on the details of the orbital shape of the original quark potential.
The $a_{c\bar c}$, $c_{c\bar c}$, and  $c_{c c}$ in \eqref{eq:Vccbar} and \eqref{eq:Vcc} are $c$-numbers. 
The size of $c_{c\bar c}$ is taken so that the term gives the observed $\Jpsi$-$\eta_c$ mass difference.
The $c_{c c}$ and $c_{\bar c\bar c}$ are taken to be equal to $c_{c\bar c}$
because there are not enough measurements of the doubly-charmed baryons to extract the value.
The spin-independent color term, $(\lambda\lambda) a_{c\bar c}$, only gives the overall energy shift of the $c\bar cc\bar c$ systems in this model.
We introduce the term to reproduce the $c\bar c$ meson masses. The scattering variables do not depend on the value of $a_{c\bar c}$.
%
The mass of the meson is written 
by the quark masses $m_c$, 
the kinetic term $\omega_0={1\over m_c \tilde b^2}$,  $a_{c\bar c}$, 
and  $C_{c\bar c}={16\over 3} c_{c\bar c}$, as
\begin{align}
M_m &= M_0 + \bra \sigma\sigma\ket C_{c\bar c}~,~~~~~
M_0=2m_c + {3\over 4}\omega_0 
-{16\over 3}a_{c\bar c}
\label{eq:mesonmass}
\end{align}%
Parameters are summarized 
in Table \ref{tbl:parameters}. 
Free parameters here are $m_c$ and $\omega_0$. We take a typical value for $m_c$ while $\omega_0$ is taken to reproduce the meson excitation energy.
\begin{table}[th]
\caption{Parameters in this model. All entries except for the size parameter $\tilde b$ are in MeV. 
Meson masses are taken from Ref.\ \cite{ParticleDataGroup:2024cfk}.
}\label{tbl:parameters}%
\begin{tabular}{ccccccccccc}
\hline
$m(\eta_c)$ & $m(\Jpsi)$ &   $ m_c $ & $a_{c\bar c}$ &  $ C_{c\bar c}(={16/3}\, c_{c\bar c})$ &$\omega_0$ &$\tilde b$(fm) \\
\hline
2984.1$\pm$0.4  & 3096.900$\pm$0.006 & 1500   & 36.34 
& 28.2 & 350 &0.385\\
\hline
\end{tabular}
\end{table}

\subsection{The resonating group method (RGM) equation}

By integrating the internal motion in the mesons, we have an equation of motion 
for the mesons, 
or the resonating group method (RGM) equation, as
\begin{align}
\int d^3r' &\sum_{\beta}\Big({\cal H}_{\alpha\beta}(r,r')-E{\cal N}_{\alpha\beta}(r,r')\Big)\chi_\beta(r') = 0
\label{eq:RGMeq}\\
{\cal O}_{\alpha\beta}(r,r') &= \int d^3\xi d^3\eta d^3\zeta\,\psi_\alpha\delta^3(\zeta-r)\,  O_q\, (1-P_{24})\psi_\beta\delta^3(\zeta-r')~,
\end{align}%
where $O_q$ is an operator for quarks, $H_q$ or 1, and  $\alpha$ or $\beta $ corresponds to the two-meson channel
in eqs.\ \eqref{eq:psiJ2}-\eqref{eq:psiJ0b}.
The $\xi$, $\eta$, and $\zeta$ are the Jacobi coordinates of the four-quark system.
For the present case, all quark masses are equal. 
Since we assume that the quark exchange only occurs in the $(0s)^4$ state,
it modifies the $0s$ part of the normalization kernel ${\cal N}$, which corresponds to $O_q=1$.
So, the normalization kernel can be expanded by the $S$-wave Harmonic oscillator wave function, $\phi_{ns}$, as
%
\begin{align}
{\cal N}_{\alpha\beta}(r,r')&= \delta^3(r-r') + \bar \nu_{\alpha\beta}\,\phi_{0s}(b,r) \phi_{0s}^*(b, r')
\\
&=(\delta_{\alpha\beta}+\bar \nu_{\alpha\beta})\phi_{0s}(b,r) \phi_{0s}^*(b, r')
+\sum_{n\ge 1}\phi_{ns}(b,r) \phi_{ns}^*(b, r')
\\
\bar \nu_{\alpha\beta}&=-\bra \alpha| P^{sfc}_{24}|\beta\ket ,~~~
b = 1/\sqrt{m_c \omega_0} = \tilde b/\sqrt{2}
\end{align}%
where $(-\bar \nu)$ is the matrix elements of the quark interchange operator in the spin-flavor-color space,
$b$ the size parameter of the relative motion between the two mesons in the orbital $(0s)^4$ configuration.
The Hamiltonian kernel can also be expanded as
\begin{align}
{\cal H}_{\alpha\beta}(r,r')&= (M_a+M_b)\calN_{\alpha\beta}(r,r') + {\cal K}_{\alpha\beta}(r,r') + {\cal V}_{\alpha\beta}(r,r')
\\
{\cal K}_{\alpha\beta}(r,r')&= \sum_{nn'} \phi_{ns}(b,R)K_{nn'}^*\phi_{n's}(b, R')\delta_{\alpha\beta}
+\bar \nu_{\alpha\beta}\sum_{n+n'\le 1}\phi_{ns}(b,r)K_{nn'}\phi_{n's}^*(b, r')
~,
\\
{\cal V}_{\alpha\beta}(r,r')&= c_{c c}\,\sum_{i\in a,j\in b} \phi_{0s}(b,r)\,
\bra \lambda_i\lambda_j \sigma_i\sigma_j (-P_{24}^{sfc})\ket 
\phi_{0s}^*(b,r')
\end{align}%
with the meson mass, $M_a$, in \eqref{eq:mesonmass}. 
The coefficients of the expansion of the kinetic term are
\begin{align}
K_{nn}&={\omega_0\over 2}(2n+{3\over 2}),~~~K_{nn+1}={\omega_0\over 2}\sqrt{(n+1)(n+{3\over 2})}~.
\end{align}%
For the $c\bar c c\bar c$, the factor $\bar \nu$ becomes 
\begin{align}
\bar \nu &= 
\left(\begin{array}{cc}
-{1\over 6} & \sqrt{1\over 12}
\\
 \sqrt{1\over 12}&{1\over 6}
\end{array} 
\right)                \text{  for $J=0$ },~~~~
\bar \nu = 
-{1\over 3} \text{  for $J=1$ and 2 }
~.
\end{align}%
%

One can also rewrite \eqref{eq:RGMeq} the RGM equation as,
\begin{align}
\Big({\cal N}^{-1/2}{\cal H}{\cal N}^{-1/2}-E\Big){\cal N}^{1/2}\chi = 0~.
\end{align}%
The above equation can be regarded as the Schr\"odinger equation with the potential defined by
\begin{align}
V(r,r') &={\cal N}^{-1/2}{\cal H}{\cal N}^{-1/2}-
\big\{(M_a+M_b)+{\cal K}\big|_{\bar \nu=0}\big\}
~~.
\end{align}
We use this $V$ as a potential between the two-meson system with the observed masses, which is
\begin{align}
V_{\alpha\beta}(r,r') &=V^K_{\alpha\beta}(r,r')+V^{CS}_{\alpha\beta}(r,r')
\label{eq:V}\\
V^K_{\alpha\beta}(r,r')&=(\sqrt{1+\bar \nu}-1)_{\alpha\beta}\,K_{01}\Big(\phi_{0s}(b, r)\phi_{1s}^*(b, r')+\phi_{1s}(b, r)\phi_{0s}^*(b, r')\Big)
\\
V^{CS}_{\alpha\beta}(r,r')&= \tilde c_{\alpha\beta} \,\phi_{0s}(b, r)\phi_{0s}^*(b, r')
\\
\tilde c &=c_{c c}\,(1+\bar \nu)^{-1/2} \Big(\sum_{i\in a,j\in b}\bra \lambda_i\lambda_j \sigma_i\sigma_j (-P_{24}^{sfc})\ket\Big) (1+\bar \nu)^{-1/2} 
\\
&= 
\left(\begin{array}{cc}
~2.443 & -0.755
\\
-0.755 &~2.057
\end{array} 
\right)C_{c c}\text{  for $J=0$,  }~~~
2C_{c c} \text{  for $J=1$,~~ and }~
0 \text{  for $J=2$ }
~.
\end{align}
%
$V^{K}$ in \eqref{eq:V} arises from the quark Pauli-blocking effect on the kinetic term.
Though we assume that the quark interchange occurs only when the system is in the $(0s)^4$ configuration,
the $0s$-$1s$ mixing term is also modified because $\Porb\phi_{0s}=\phi_{0s}$. 
The $0s$-$1s$ component becomes larger when $\bar \nu>0$, as in the $(\Jpsi\Jpsi)|_{J=0}$.
This enhanced mixing causes an attraction between the mesons in the energy range below about $\omega_0$.
This attraction comes from the many-body effect of quarks; 
it arises because the final two-meson state 
comes from the initial two-meson state with and without the quark rearrangement.
On the other hand, the system has a repulsion when $\bar \nu < 0$.
In the $c\bar c c\bar c$ systems, all the channels except for the $(\Jpsi\Jpsi)|_{J=0}$ are repulsive.
Namely, 
the Pauli-blocking effect is dominant in the $c\bar c c\bar c$ systems.
As for the $V^{CS}$, the effects are repulsive or zero in all the channels. 
The hadron interaction that originated from the quark degrees of freedom 
is repulsive in the $c\bar cc\bar c$ systems in the low-energy region.

\section{Results}

\subsection{Potential}

\begin{figure}[htbp]
\begin{center}
\includegraphics[clip,  scale=0.35]{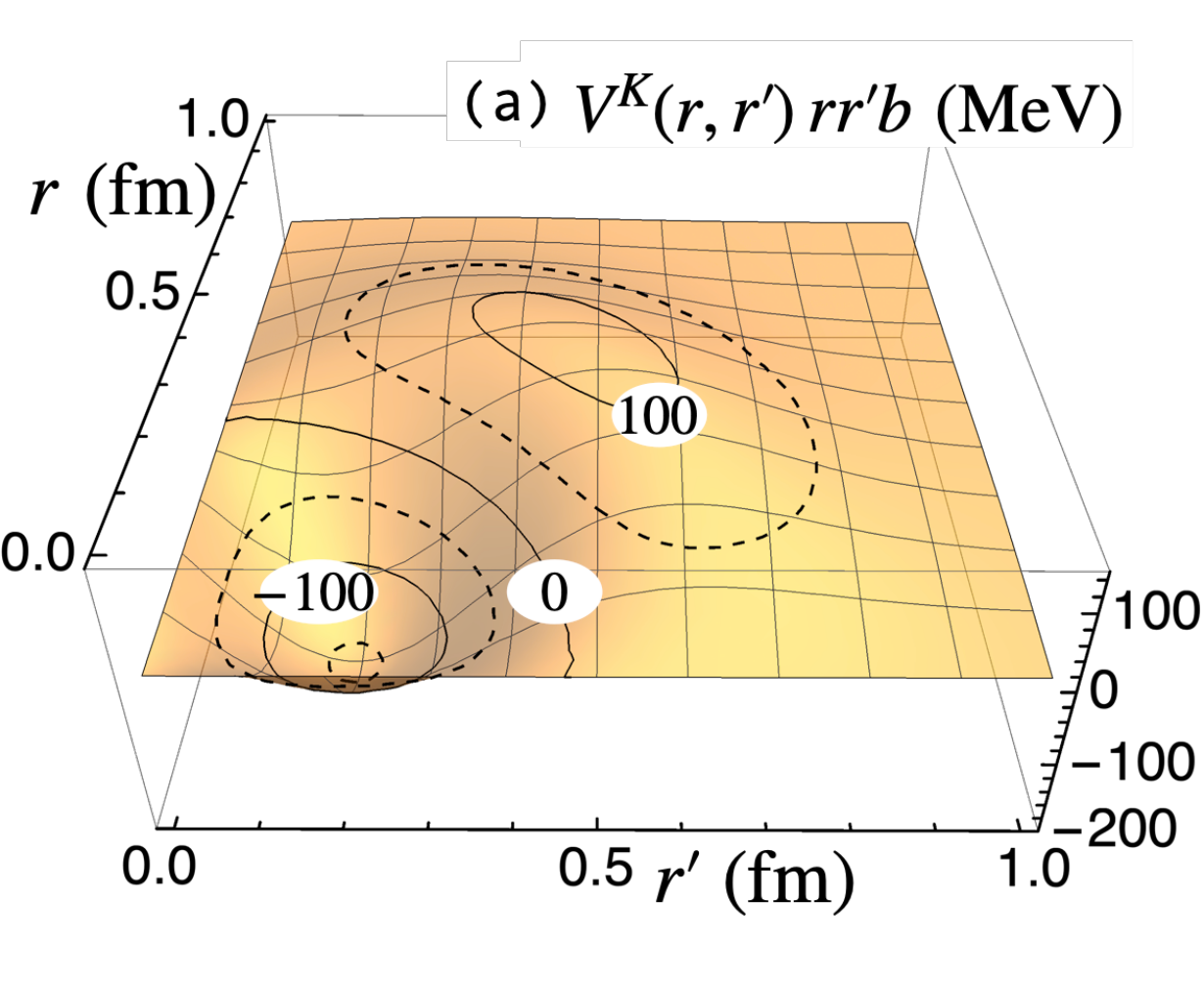}\hfill
\includegraphics[clip,  scale=0.37]{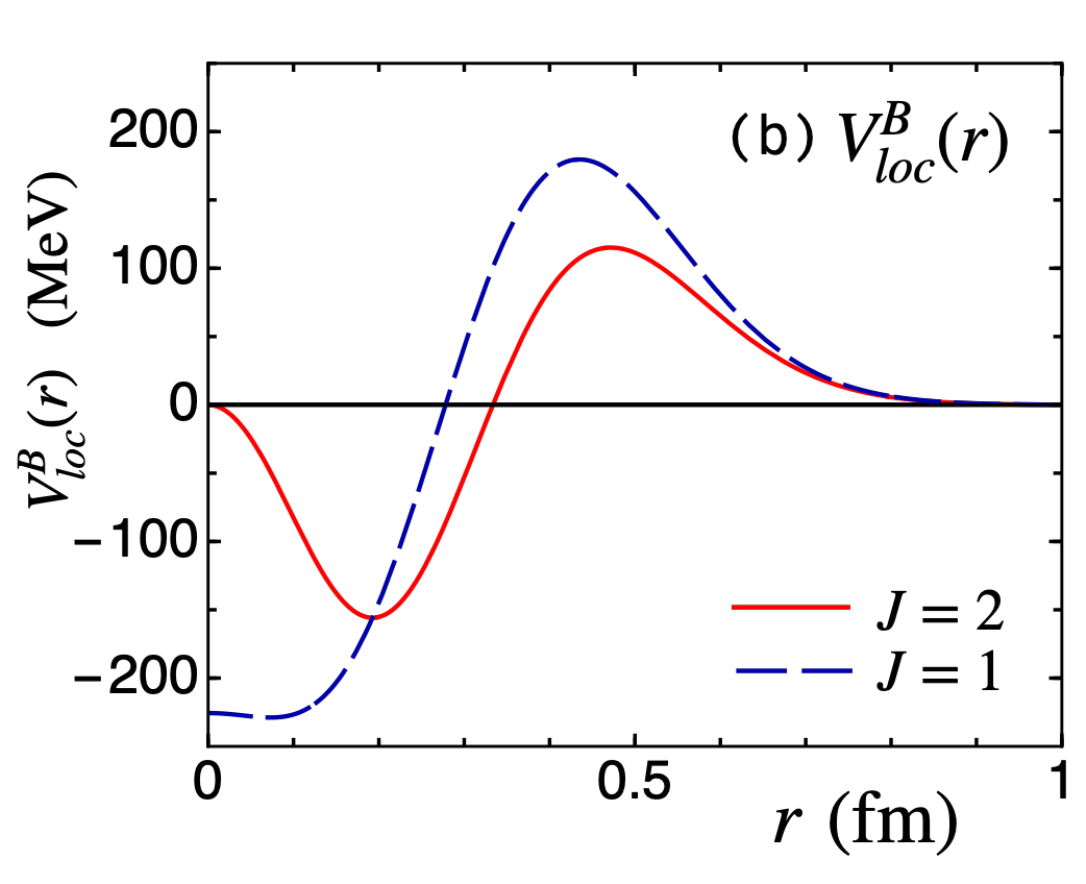}
\caption{(a) The nonlocal potential for $J=2$, (b) the local potential that gives the same phase shift by Born approximation as the nonlocal potential for $J=1,2$. }
\label{fig1}
\end{center}
\end{figure}

The obtained nonlocal potential for the $c\bar cc\bar c$ $J=2$ channel is shown in Figure \ref{fig1}(a).
The potential corresponds to $V_K$ 
because the effect of the color spin term, $V^{CS}$, vanishes in the $J=2$ channel. 
It has a repulsive barrier in the middle range ($\sim 0.5$ fm), 
so the system has repulsion in the low-energy region.
In the very short range, it has a negative dip. 
It is known that this kind of well-and-barrier shape of the potential can cause a shape resonance in the scattering.

We use the nonlocal potential directly in our calculation.
However, to understand the situation more intuitively,
let us show the local potential for the single channel, $V^B_{loc}(r)$,  in Figure \ref{fig1}(b).
$V^B_{loc}(r)$ gives the same phase shift 
as the nonlocal potential gives by the Born approximation,
\begin{align}
\tan\delta^B(k)&=-k\int_0^\infty r^2 \rmd rr'{}^2 \rmd r'\, j_0(kr) 2\mu V(r,r') j_0(kr')
 ~.
\end{align}%
The analytic form of the local potential can be obtained as
\begin{align}
 V_{loc}^B(r) &=2\int_{-\infty}^\infty \rmd z (r^2-z^2)V_{nonloc}(r+z,r-z)
\\
 &= 12 \omega_0(\sqrt{1+\bar \nu}-1)_{\alpha\beta}
 \Big(1-{2r^2\over 3b^2}\Big){r^2\over b^2}\exp[-{r^2\over b^2}]
 +4\tilde c_{\alpha\beta} \, \Big(1-{r^2\over 2b^2}\Big)\exp[-{r^2\over b^2}]
 ~.
\label{eq:eqivLocPot}
\end{align}%
The above local potential slightly deviates from the phase-shift equivalent local potential obtained by solving the inverse scattering problem numerically.
Nevertheless, we show this because the feature is more clearly seen.
It has a node $ r\sim \sqrt{3/2}\, b$, 
which corresponds to the zero point of the scattering phase shift at $E \sim 3\omega_0/4$. 
The numerical shape for the $J=1,2$ channels in Figure \ref{fig1}(b)
shows that there is a barrier in the potential, which causes a rapid increase in the phase shift.

\subsection{Phase shifts and cross section}

\begin{figure}[htbp]
\begin{center}
\raisebox{1.95cm}{\includegraphics[clip,  scale=0.33]{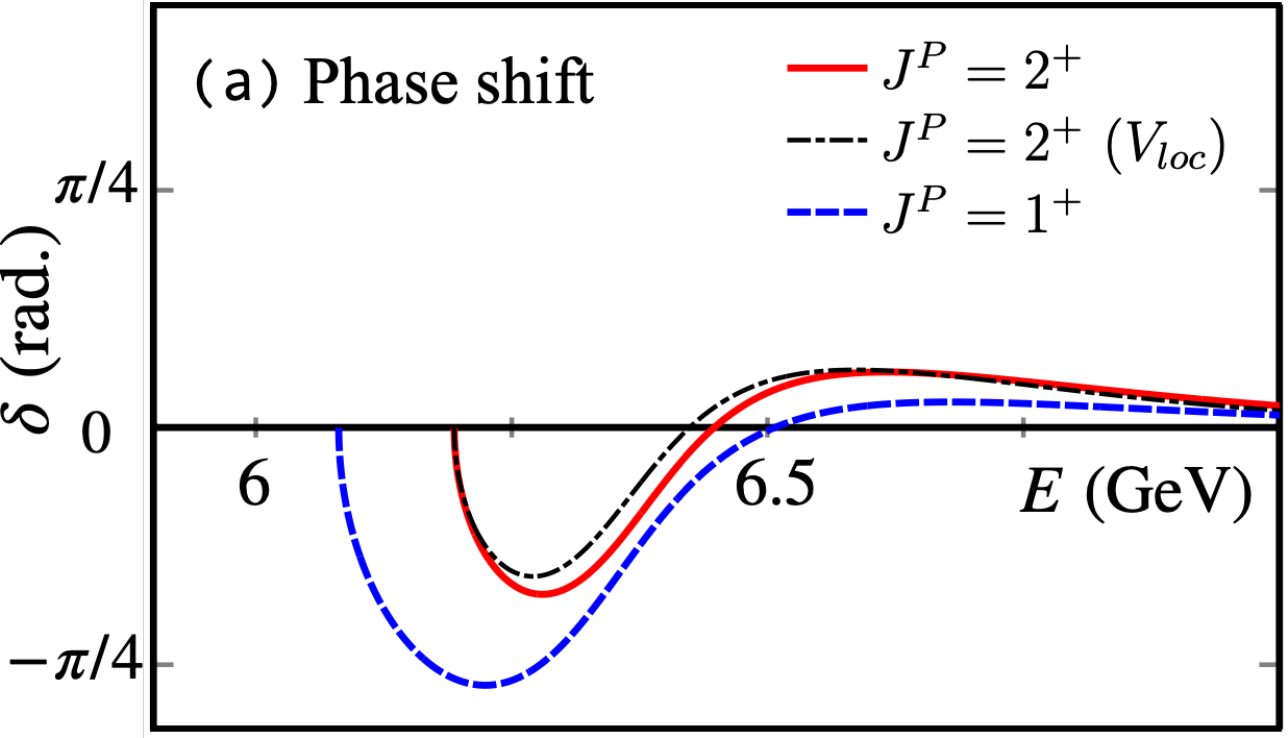}}\hfill
\includegraphics[clip,  scale=0.33]{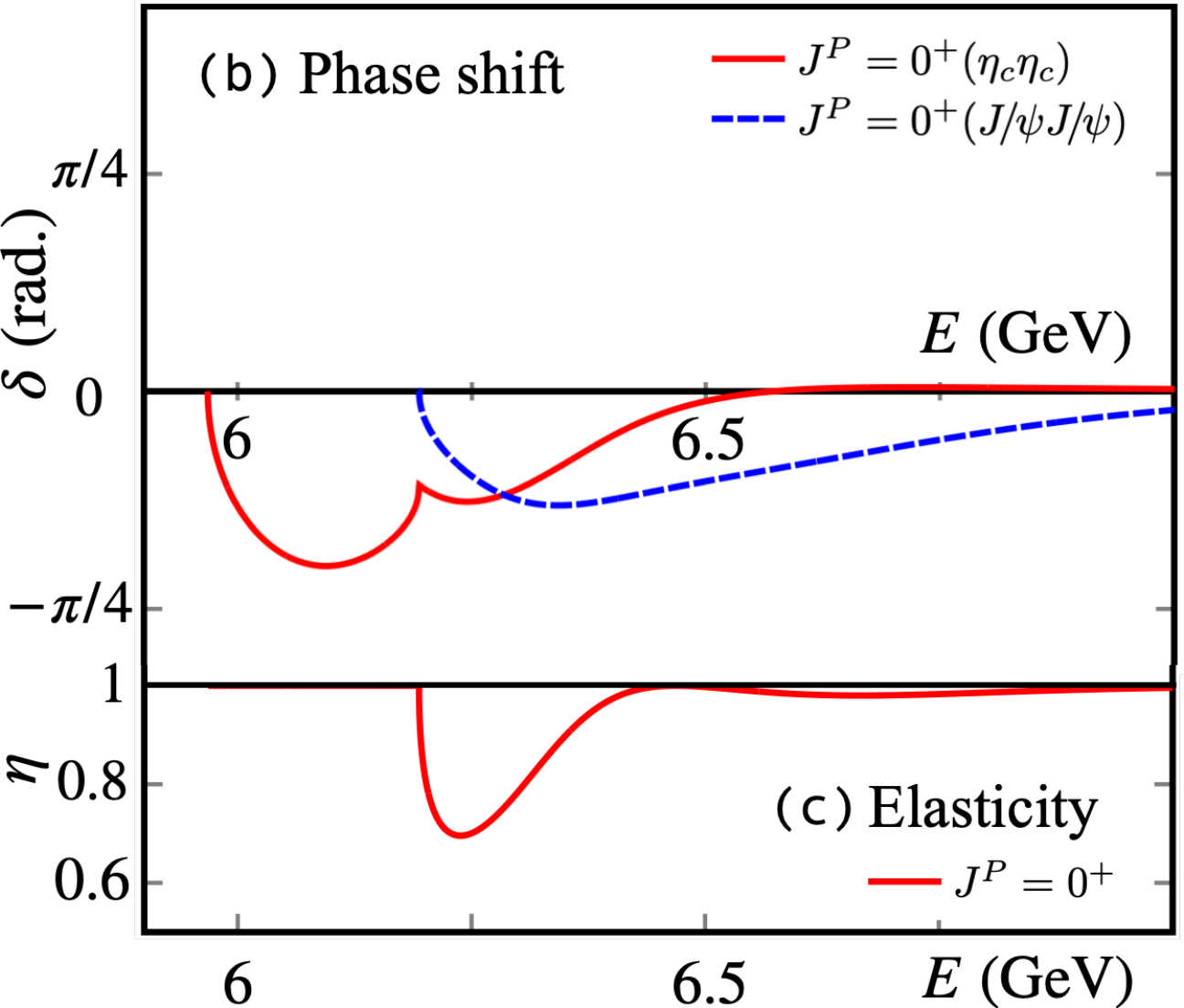}\hfill
\caption{
(a) The scattering phase shifts for $J=1$ and 2. That with the local potential is also shown for $J=2$.
(b) The scattering phase shifts and (c) the elasticity $\eta$ for $J=0$.}
\label{fig2}
\end{center}
\end{figure}

\begin{figure}[htbp]
\begin{center}
\includegraphics[clip,  scale=0.32]{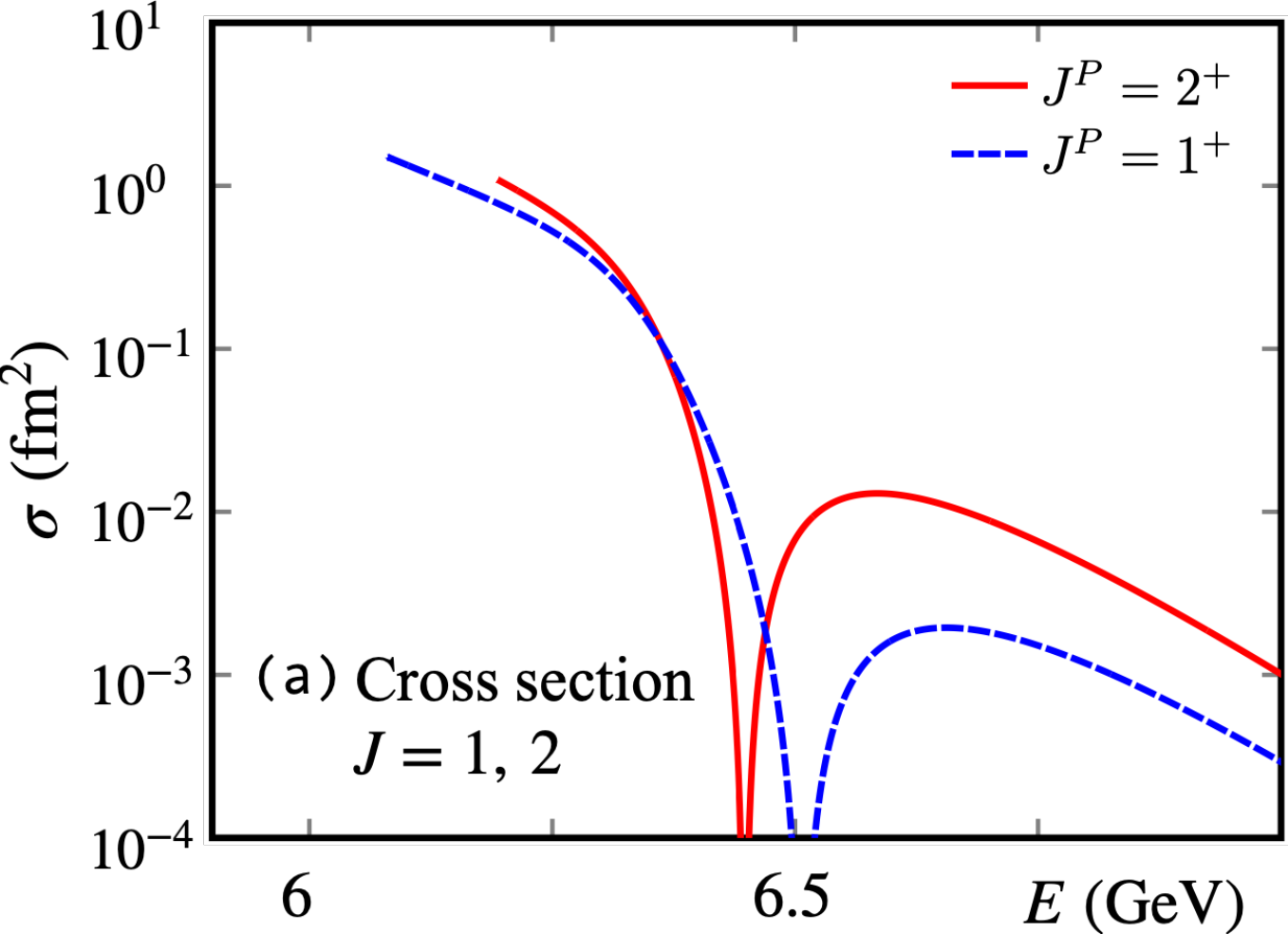}\hfill
\includegraphics[clip,  scale=0.32]{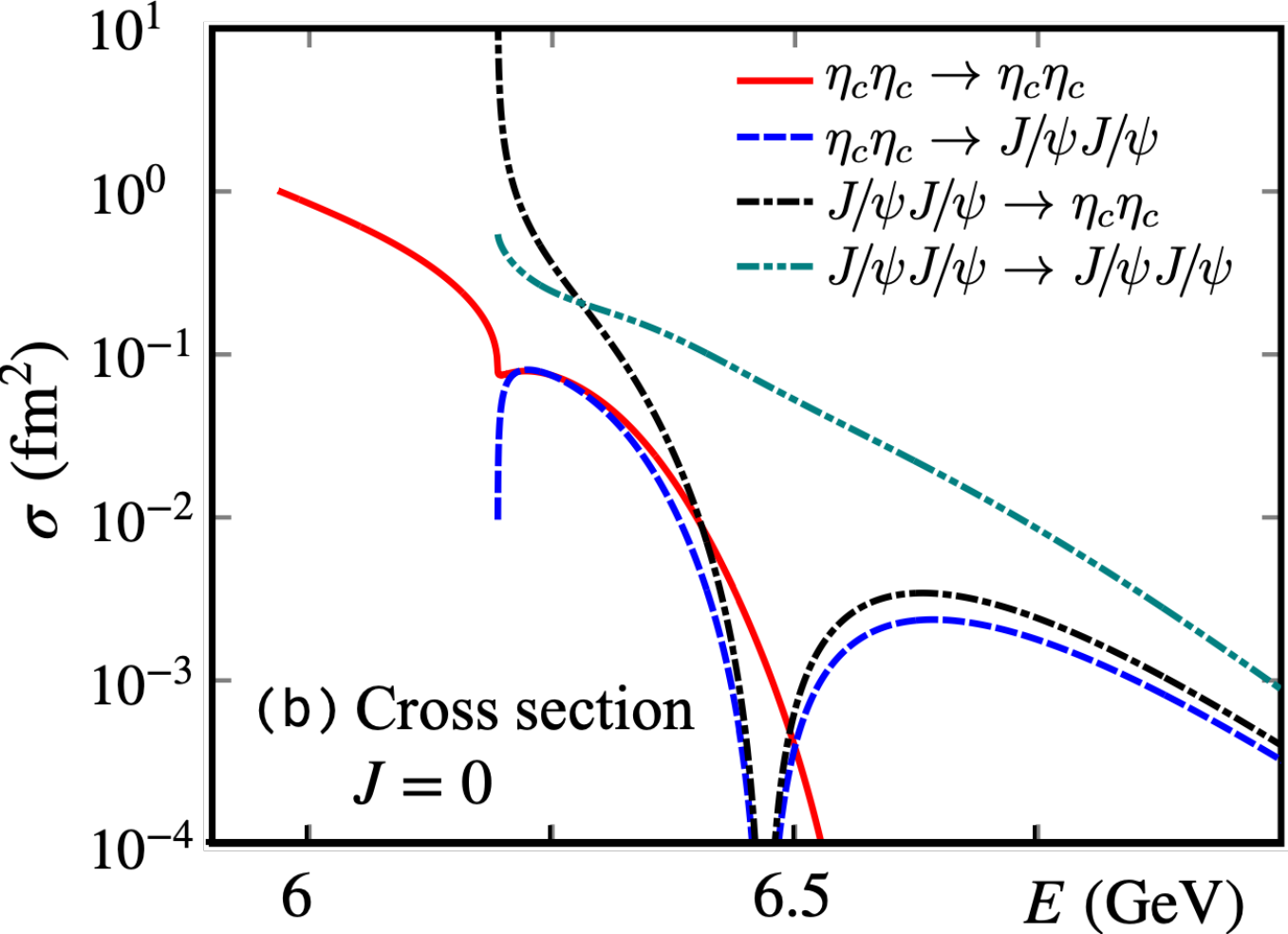}
\caption{(a) The cross sections for $J=1$ and 2 and 
(b) for $J=0$.}
\label{fig3}
\end{center}
\end{figure}

The phase shifts and the cross-sections of the two-$c\bar c$ meson scattering are shown in Figures \ref{fig2} and \ref{fig3} for each of the $J=0,1,2$ channels.
The component of the $J=2$ channel is $\Jpsi\Jpsi$, whereas that of the $J=1$ is $\Jpsi\eta_c$.
The $J=0$ has two channels,  $\eta_c\eta_c$ and $\Jpsi\Jpsi$.

The $J=2$ 
phase shift is shown in Figure \ref{fig2}(a).
Since the spin-color term vanishes in this channel,
the quark antisymmetrization effect, namely, the potential $V_K$, 
alone gives the scattering phase shift shown here.
The phase shift has a node at ${4\over 3}\omega_0$ above the threshold, around 6.5 GeV.
The result from the local potential \eqref{eq:eqivLocPot} is also shown in the figure,
which is essentially the same as that of the nonlocal one.
So, this rapid increase of the phase shift at around 6.3-6.4 GeV 
comes from the potential's well-and-barrier shape.
The increase is not large enough to be regarded as a resonance, 
but if it is seen experimentally, that is very likely the quark many-body effect.
The cross-section of the $J=2$ channel (Figures \ref{fig3}(a)) has a dip at that energy.

The $J=1$ channel gets repulsion from $V_{CS}$ as well as $V_K$.  
Due to this extra repulsion, the node in the phase shift and the dip in the 
cross section are pushed upward.
However, the node energy itself is similar to that of the $J=2$ channel because the threshold is lower.
Observing the $\Jpsi\eta_c$ mode experimentally may be more challenging, 
but the quark effect will be seen at around 6.2-6.3 MeV.

As for the $J=0$, 
the $V_K$ is repulsive in $\eta_c\eta_c$ and slightly attractive in $\Jpsi\Jpsi$.
The $V_{CS}$ is repulsive in both of the channels.
As a result, both of the two channels get repulsion, as shown in Figure \ref{fig2}(b).
The node almost disappears in $\eta_c\eta_c$ and cannot be seen in $\Jpsi\Jpsi$.
Their mixing of the two channels is rather small, as seen in Figure \ref{fig2}(c).

This quark Pauli-blocking effect cannot be easily seen if one only examines the system using the bound state approach.
The potential from the kinetic term, $V_K$, disappears in the bound state 
if its orbital configuration is $(0s)^4$.
This may be why the effect has not been investigated in detail so far.
On the other hand, this effect has already been included in the RGM scattering calculation. For example, 
some of the resonances that appear below $\eta_c\eta_c'$ threshold in \cite{Ortega:2023pmr} may correspond to the 
rapid increase in the phase shift in the present work.
Also, we point out that the quark Pauli-blocking effect strongly depends on the channels.
That is a different feature 
from the work \cite{Wang:2023jqs}, in which they introduced a new type of gluon confinement configuration
to explain the states below 7 GeV.

The di-$\Jpsi$, or $c \bar c c\bar c$ systems are
almost free from the pion interaction and it is the place  where one can see the quark effect directly.
We hope to have more experimental results in the near future.
\bigskip

This work is also  supported by the RCNP Collaboration Research Network program as the project number COREnet-2024 (project 54)  by (ST).

\bibliographystyle{unsrt}
\bibliography{QCHSC24_STakeuchi}

%

\end{document}